\documentclass{acm_proc_article-sp}
\usepackage{graphicx}
\usepackage{balance} 
\usepackage{multirow}
\usepackage{listings}
\usepackage{url}
\usepackage[usenames,dvipsnames]{color}
\usepackage{pgfplots}
\begin{document}

\title{decimalInfinite: All Decimals In Bits.\\ No Loss. Same Order. Simple.}

\numberofauthors{1}

\author{
\alignauthor
Ghislain Fourny\\
       \affaddr{28msec, Inc.}\\
       \affaddr{Z\"urich, Switzerland}\\
       \email{g@28.io}
}
\date{6 March 2015}
\maketitle

\begin{abstract}
This paper introduces a binary encoding that supports arbitrarily large, small and precise decimals. It completely preserves information and order. It does not rely on any arbitrary use-case-based choice of calibration and is readily implementable and usable, as is. Finally, it is also simple to explain and understand.
\end{abstract}

\section{introduction}
In the early stages of computers, storage was scarce. And it was even scarcer in the processor itself. The first computers supported basic arithmetic operations on small integers (e.g., 4 bits). As more memory was needed, the size of the registers in the processor architecture was increased to finally reach the current 64 bits.

While this was mostly driven by the size of the memory, which never failed to exceed the maximum value space at some point, this also had an impact on the size of integers and decimals processed (e.g., Von-Neumann \cite{VonNeumann1946}).

The numbers used in programs are of two main kinds: integers and decimals. Integers are typically bounded to some range. In the case of decimals, it is a bit more complicated. They are limited both in their range (driven by the size of their exponent) and precision.

For scientists that need to use bigger numbers and/or more precision, special libraries are available, and domain-specific software \cite{MAPLE} \cite{MATHEMATICA} \cite{MATLAB} make it possible to overcome the limitations of processors.

In the modern database era, notably document stores, syntaxes such as XML \cite{XML} and JSON \cite{JSON}, but also their data models and the associated query languages, do not impose any limitation on integers or decimals: on the logical level, the entire value space is covered and the limit is only the size of the available storage.

In the context of document stores, decimal numbers need to be efficiently stored and retrieved, but they also need to be used as index values. Indexing documents on decimal values is the primary use case for decimalInfinite. For efficiency reasons, it is desirable that an index lookup can be done without decoding the decimals stored in a hash index (point queries) or a tree index (range queries). In the latter case, this is possible when the (lexicographic) order of encoded values corresponds to the ordering of the corresponding numbers. Then, only the queried decimals need to be encoded once, and the remaining comparisons occur on the binary sequences only. 

When decimalInfinite was designed, to the extent of our knowledge and our investigations, state-of-the-art databases did not support yet such an encoding that would cover the entire decimal space (that is, the entire value space of the XML xs:decimal type, or the entire value space of JSON numbers). However, many encodings exist that have some of the desired properties. decimalInfinite unifies the ideas in these encodings in such a way that all these properties apply simultaneously. In particular, none of the separate schemes used are new: decimalInfinite is only original in the way it puts them all together.

This paper contributes an encoding that solves this problem. More recently, we became aware of the Rishe encoding (Section \ref{section-rishe}), which solves the same problem. However, decimalInfinite uses a different approach and does not require any calibration of choice of intervals. decimalInfinite also improves the asymptotic complexity for large decimals or integers.

The encoding and decoding algorithms have been implemented in C++ and are used on production machines to store decimals as BSON binaries \cite{BSON}, on a MongoDB \cite{MongoDB} data store. Another implementation is available in JSONiq \cite{decimalInfiniteJSONiq}: the encoding takes only 117 lines of code.

The decimalInfinite encoding relies on decomposing the underlying decimal into a normal form, as is typically done in other encodings. This decomposition is made of an overall sign, an exponent sign, an exponent and a significand (also called mantissa by Von Neumann, although, as is common practice, we prefer to avoid this terminology because of its different meaning for logarithm computation). These four components are encoded in turn and in this order.

Section \ref{section-related-work} gives an overview of the integer and decimal encoding landscape.
Section \ref{section-ordering} gives an overview of the main two ways of sorting sequences of bits.
Section \ref{section-encoding} gives the algorithm for encoding decimals in the decimalInfinite format.
Section \ref{section-decoding} gives the decoding algorithm.
Section \ref{section-proof} gives a detailed proof of the main property of decimalInfinite (that it is order preserving).
Section \ref{section-examples} shows the encoding of the smallest integers.
Section \ref{section-fine-tune} shows how the core of the encoding can be extended to special numbers (NaN, infinity, etc).
Section \ref{section-complexity} briefly summarizes complexity aspects, including a comparison to the Rishe encoding.
Section \ref{section-implementation} gives a few implementation details.

\section{Related work}

There are various encodings of integers and decimals commonly used in practice and found in literature. In general, one needs to distinguish between schemes for encoding digits and integers on the one hand (possibly used to encode decimals with fix-point semantics), and schemes based on the former that support floating point encodings on the other hand.

\label{section-related-work}

\subsection{Natural encoding}

Positive integers have a natural encoding which is simply their representation in base 2, as shown on Figure \ref{figure-natural-encoding}.

\begin{figure}[p]
\caption{Natural encoding of an integer (base 2), not padded and padded to 4 bits.}
\label{figure-natural-encoding}
\center
\begin{tabular}{|l|l|l|}
\hline
Integer & Binary representation & Padded representation \\
\hline
1 & 1 & 0001 \\
\hline
2 & 10 & 0010 \\
\hline
3 & 11 & 0011\\
\hline
4 & 100  & 0100 \\
\hline
5 & 101  & 0101\\
\hline
... & ... & ... \\
\hline
\end{tabular}
\end{figure}

It supports an unlimited range, however is not order-preser\-ving.

If however the range is limited, for example to 8, 16, 32 or 64 bits, as is commonly done in most programming languages, and the encodings are padded with leading 0s, then this encoding becomes order-preserving.

\subsection{Signed Integers}

Signed integers are commonly stored by encoding the sign in the first bit (Figure \ref{figure-natural-signed-encoding}), meaning that positive integers (beginning with a 0, which is half the range) are stored in the same way as the unsigned encoding, while negative integers are stored beginning with a 1. Lexicographic order is only preserved for positive integers, as well as for negative integers, but not overall.

\begin{figure}[p]
\caption{Natural encoding of signed integers (base 2, padded to 4 bits).}
\label{figure-natural-signed-encoding}
\center
\begin{tabular}{|l|l|}
\hline
Integer & Signed binary representation \\
\hline
-3 & 1101 \\
\hline
-2 & 1110 \\
\hline
-1 & 1111 \\
\hline
0 & 0000 \\
\hline
1 & 0001 \\
\hline
2 & 0010 \\
\hline
3 & 0011 \\
\hline
... & ... \\
\hline
\end{tabular}
\end{figure}

\subsection{Elias Gamma Code}
\label{section-gamma-code}

Gamma codes are a variable-length encoding that supports the entire non-negative integer range ($\mathbb{N}$). One of their main usage and motivation is that they are prefix codes, meaning that they can get concatenated to each other in a way that they can still be separated again unambiguously. Figure \ref{figure-gamma-encoding} shows how the Gamma code looks like for the first integers.

\begin{figure}[p]
\caption{The Gamma Code for the first non-negative integers, and the modified, order-preserving gamma code.}
\label{figure-gamma-encoding}
\center
\begin{tabular}{|l|l|l|l|}
\hline
Integer & Offset by 1 & Gamma & Modified Gamma \\
\hline
0 & 1 (1) & 1 & 0\\
\hline
1 & 2 (10) & 01 0 & 10 0 \\
\hline
2 & 3 (11) & 01 1  & 10 1\\
\hline
3 & 4 (100) & 001 00 & 10 00\\
\hline
4 & 5 (101) & 001 01 & 110 01\\
\hline
5 & 6 (110) & 001 10 & 110 10\\
\hline
... & ... & ... & ... \\
\hline
\end{tabular}
\end{figure}

 The main idea is that a first sequence of 0s, terminated by a 1, encodes the length of the binary-representation of the integer. Then, the natural binary representation of the integer (most of the time offset by 1), without its leading 1, follows.

Since the initial number of 0s is identical to the number of digits that follow after the 1, it is possible to unambiguously deduct where the encoding stops, solely relying on the number of leading 0s.

For example, 00110010 can be unambiguously separated into 00110 (5) and 010 (1).

The Gamma code in its original form is not order preserving. However, simply inverting the first sequence of 0s and the terminating 1 solves the issue, as is shown in the last column of Figure \ref{figure-gamma-encoding}.

\subsection{BCD, Chen-Ho, DPD}

Regardless of whether decimals are stored in fixed point format or in floating point format, the sequence of their significant digits needs to be encoded.

The Binary-coded decimal encoding (BCD) encodes each digit in a group of 4 or 8 digits, so-called tetrades. Improvements include the Chen-Ho \cite{ChenHo} encoding, which manages to encode 3 digits on 10 bits (declets). It has the nice particularity of being extremely efficient to process (no multiplications, no divisions), and of being friendly to decimal computations. However, it does not preserve order. The Densely packed decimal (DPD) encoding \cite{DPD} is an improvement upon the Chen-Ho encoding.

\subsection{IEEE float and double encoding}

The IEEE 754 standard specifies a couple of standard, floating-point encoding for decimals more commonly known as float or double in mainstream programming languages. It both supports a finite range of decimals (its length is fixed), and is not order-preserving. It has several variants (binary16, binary32, binary64, binary128, decimal32, decimal64, decimal128) depending on the length and on the way the significand is encoded. These encodings rely on DPD.

\subsection{IBM Patent}
The US Patent 7685214 (``Order-preserving encoding formats of floating-point decimal numbers for efficient value comparison'') , filed by IBM, solves the order-preserving issue, but with a finite-length encoding, which implies that it supports a finite range of decimals only.

One very interesting idea in this approach is that, if the sign is negative, the significand $m$ is encoded as $10-m$ to preserve the order.

\subsection{The Matula-Kornerup and the Rishe Encodings}
\label{section-rishe}

The Rishe Encoding \cite{Rishe1992} is the closest match to decimalInfinite we found in literature in terms of problem solving. It supports arbitrarily large, small and precise decimals, is compact, and is also compatible with a bitwise lexicographic comparison. However, it relies on an arbitrary choice of intervals (128) based on the use case. decimalInfinite does not rely on such a choice and scales up continuously, regardless of how large, small or precise decimals are.

The Rishe encoding was itself proposed as an improvement upon the Matula-Kornerup encoding \cite{Kornerup1983}, which relies on a representation of the decimal as a continuous fraction. The latter did not scale up with exponents. The exponent part of Rishe scales up logarithmically in the decimal, while that of decimalInfinite scales up double-logarithmically (see Section \ref{section-complexity}).

\section{Ordering sequences of bits}
\label{section-ordering}

There are two widespread ways of sorting sequences of bits, as shown on Figure \ref{figure-orders}. The first one, pseudo-lexicographic (also called in literature shortlex, quasi-lexicographic, length-lexicographic), first orders by size, and then within a size, lexicographically. The second one, regardless of the size, compares the bits from the left to the right, with the convention that, when a sequence is a prefix of another, the shorter one comes first.

\begin{figure}[p]
\caption{Two main ways of sorting binary sequences}
\label{figure-orders}
\center
\begin{tabular}{|c|l|}
\hline
Pseudo-lexicographic order & Full lexicographic order\\
\hline
$0$ & $0$ \\
$1$ & $00$ \\
$00$ & $000$ \\
$01$ & $001$ \\
$10$ & $01$ \\
$11$ & $010$ \\
$000$ & $011$ \\
$001$ & $1$ \\
$010$ & $10$ \\
$011$ & $100$ \\
$100$ & $101$ \\
$101$ & $11$ \\
$110$ & $110$ \\
$111$ & $111$ \\
\hline
\end{tabular}
\end{figure}

MongoDB sorts binaries pseudo-lexicographically, even if we could find no documentation regarding this. decimalInfinite preserves the ordering of decimals with the semantic understanding of the full lexicographic order.

\section{Encoding}
\label{section-encoding}
Let us now get into the details of the encoding itself. The general idea is any non-zero decimal can be expressed in a canonical scientific form with four components (sign, exponent sign, exponent, significand). These four components can be encoded separately and concatenated. Since each of the components (but the last one) is a prefix code, it can be unambiguously decoded again.

\subsection{Canonical decomposition}

Zero is handled separately and encoded as $10$. Any non-zero decimal number can be expressed uniquely in scientific notation as in commonly done in literature, that is, in the form $$s\times m \times10^{t\times e}$$ where:

\begin{itemize}
\item The overall sign is $s\in \{-1, 1\}$.
\item The exponent $e\in \mathbb{N}$ is a non-negative integer (which is the absolute value of the logarithm in base ten of the absolute value of the original number, rounded down to the next integer).
\item The exponent sign is $t\in \{-1, 1\}$.
\item The significand is $m\in [1,10)$, a real number between 1 (included) and 10 (excluded).
\end{itemize}

If S denotes the encoding s, T that of t and so on, then the overall encoding comes naturally as STEM as shown on figure \ref{figure-overall-encoding}. This is because decimal numbers in scientific notation can be sorted with the following criteria in this order:
\begin{enumerate}
\item sign
\item exponent sign
\item exponent
\item significand
\end{enumerate}

\begin{figure}[p]
\caption{Encoding of an overall decimal in scientific notation $s\times m \times10^{t\times e}$.}
\label{figure-overall-encoding}
\center
\begin{tabular}{|l|l|l|l|}
\hline
$S$ & $T$ & $E$ & $M$ \\
\hline
\end{tabular}
\end{figure}

Throughout this paper, four examples, which cover various combinations of the four components, will be used:

$$-103.2 = - 1.032 \times 10^2$$
$$-0.0405 = -4.05 \times 10^{-2}$$
$$0.707106 = 7.07106 \times 10^{-1}$$
$$4005012345 = 4.005012345 \times 10^9$$

\subsection{Encoding the sign}

The sign of a decimal is encoded on two bits as shown on Figure \ref{figure-sign}

\begin{figure}[p]
\caption{Encoding of the overall decimal sign.}
\label{figure-sign}
\center
\begin{tabular}{|l|l|}
\hline
S & Sign s \\
\hline
00 &  negative sign ($s=-1$, e.g., $-4.3\times10^3$)\\
\hline
10 & positive sign ($s=1$, e.g., $4.3\times10^3$)\\
\hline
\end{tabular}
\end{figure}

Since zero is simply encoded with $10$ with no further bits, it is already apparent that its encoding appears lexicographically after the encoding of any negative decimal, and before the encoding of any positive decimal.

The reason for using two bits rather than just one is that negative infinity (-INF), positive infinity (+INF) as well as negative zero and NaN can be conveniently encoded as well (see Section \ref{section-fine-tune}).

So far, our four examples have an encoding that begins as follow:

\begin{tabular}{l|l}
$- 1.032 \times 10^2$ & 00... \\

$-4.05 \times 10^{-2}$ & 00... \\

$0$ & 10 \\

$7.07106 \times 10^{-1}$ & 10... \\

$4.005012345 \times 10^9$ & 10...\\
\end{tabular}
\subsection{Encoding the exponent sign}

The exponent sign is encoded on the third bit, as shown on figure \ref{figure-exponent-sign}.

\begin{figure}[p]
\caption{Encoding of the exponent sign.}
\label{figure-exponent-sign}
\center
\begin{tabular}{|l|l|}
\hline
S and T & s and t \\
\hline
00 0 &  negative sign, non-negative exponent sign\\
\hline
00 1 & negative sign, negative exponent sign\\
\hline
10 0 & positive sign, negative exponent sign\\
\hline
10 1 & positive sign, non-negative exponent sign\\
\hline
\end{tabular}
\end{figure}

The encoding of our four examples continues as follows:

\begin{tabular}{l|l}
$- 1.032 \times 10^2$ & 00 0... \\

$-4.05 \times 10^{-2}$ & 00 1... \\

$7.07106 \times 10^{-1}$ & 10 0... \\

$4.005012345 \times 10^9$ & 10 1...\\
\end{tabular}

\vspace{10pt}
\subsection{Encoding the exponent}

The absolute value of the exponent is encoded with a modified gamma code (as explained in section \ref{section-gamma-code}), using an offset of 2 \footnote{The offset by 2 is needed, because the initial, length-discriminating sequence of the Gamma code must occupy at least two bits. With only one bit, it would be impossible to both deduce the sign of the exponent and the length of the exponent encoding.}.

\label{section-exponent-encoding}
\begin{enumerate}
\item The exponent is offset by +2, for example, 4 is encoded with the modified gamma code of 6. 
\item  The offset exponent is written in a binary form, for example, 6 is written 110.
\item  Call N the number of its digits (in the case of 110: 3).
\item  The first digit is replaced with N-1 ones, followed by a zero (in the case of 110: 110 10)
\end{enumerate}

Figure \ref{figure-exponent-encoding} shows how the smallest absolute values of the exponent are encoded.

Once the absolute value of the exponent has been encoded as shown above, it is either negated if $T=0$, or left unchanged if $T=1$. E is then obtained by dropping the first bit of the obtained string (because this first bit is already encoded in T). In other words, the (negated or not) absolute value of the exponent is encoded as TE.

Note that 0 is treated as a non-negative exponent, so that it will always be encoded as 100 if the decimal is positive, and as 011 if the decimal is negative. This means that a decimalInfinite encoding will never begin with 10011 or 00100\footnote{This could have been avoided, but at the cost of offsetting negative exponent encodings (and only them) by 1 instead of 2, which would have introduced more complexity}.

\begin{figure}
\caption{Encoding the exponent.}
\label{figure-exponent-encoding}
\center
\begin{tabular}{|l|l|l|l|}
\hline
e & e offset by 2 & TE & TE\\
  &  & (non negated, $T=1$) & (negated, $T=0$)\\
\hline
0 & 2 (10) & 10 0 & 01 1  \\
\hline
1 & 3 (11) & 10 1  & 01 0\\
\hline
2 & 4 (100) & 110 00  & 001 11\\
\hline
3 & 5 (101) & 110 01 & 001 10\\
\hline
4 & 6 (110) & 110 10 & 001 01\\
\hline
5 & 7 (111) & 110 11 & 001 00\\
\hline
6 & 8 (1000) & 1110 000 & 0001 111\\
\hline
7 & 9 (1001) & 1110 001 & 0001 110\\
\hline
8 & 10 (1010) & 1110 010 & 0001 101\\
\hline
9 & 11 (1011) & 1110 011 & 0001 100\\
\hline
... & ... & ...& ...\\
\hline
\end{tabular}
\end{figure}

The encoding of our four examples continues as follows:

\begin{tabular}{l|l}
$- 1.032 \times 10^2$ ($e=2$, opposite signs) & 00 001 11... \\

$-4.05 \times 10^{-2}$ ($e=2$, same sign) & 00 110 00... \\

$7.07106 \times 10^{-1}$ ($e=1$, opposite signs) & 10 01 0... \\

$4.005012345 \times 10^9$ ($e=9$, same sign) & 10 1110 011...\\
\end{tabular}

\subsection{Encoding the significand}

The significand is encoded in a way similar to decimal32, decimal64 and decimal128, that is:

\begin{itemize}
\item its initial digit (before the decimal point) is encoded on 4 bits (tetrade) in its natural binary representation.
\item the remaining digits (after the decimal point) are organized in groups of 3 (declets). Each declet is encoded in its natural binary representation on 10 bits. Trailing 0s are added to make sure that the last group also has 3 digits.
\end{itemize}

If the overall sign of the decimal is negative though, a trick similar to the IBM patent is used: $10-m$ is encoded instead of $m$ (in this case, the leading digit may be a 0).

\begin{figure}
\caption{Examples of significand encodings}
\label{figure-significand-encoding}
\center
\begin{tabular}{|l|l|l|l}
\hline
$8.968 (=10-1.032)$ & 8 968 & 1000 \\
& & 1111001000 \\
\hline
$5.95 (=10-4.05)$ & 5 950 & 0101 \\
& & 1110110110\\
\hline
$7.07106$ & 7 071 060 & 0111 \\
& & 0001000111\\
& & 0001111000\\
\hline
$4.005012345$ & 4 005 012 345 & 0100 \\
& & 0000000101\\
& & 0000001100\\
& & 0101011001\\
\hline
\end{tabular}
\end{figure}

The encoding of our four examples can now be completed:

\begin{tabular}{l}
$- 1.032 \times 10^2$ ($10-m$ is taken)\\
00 001 11 1000 1111001000\\
\\
$-4.05 \times 10^{-2}$ ($10-m$ is taken)\\
00 110 00 0101 1110110110\\
\\
$7.07106 \times 10^{-1}$\\
10 01 0 0111 0001000111 0001111000\\
\\
$4.005012345 \times 10^9$\\
10 1110 011 0100 0000000101 0000001100 0101011001\\
\end{tabular}

\section{Decoding}
\label{section-decoding}

Decoding is also performed from left to right, in a way similar to encoding.

\vspace{20pt}
\subsection{Decoding the overall sign}

The overall decimal sign is obtained straightforwardly from the first two bits. If no more bits follow, it is a zero. Otherwise, decoding continues with the exponent.

\subsection{Decoding the exponent}

The exponent sign can be deduced from the third bit, but depends on the overall sign:

\begin{itemize}
\item If the overall sign is - and the third bit is a 0, the exponent sign is +.
\item If the overall sign is - and the third bit is a 1, the exponent sign is -.
\item If the overall sign is + and the third bit is a 0, the exponent sign is -.
\item If the overall sign is + and the third bit is a 1, the exponent sign is +.
\end{itemize}

The exponent encoding, starting at the third bit, is of variable length. Since gamma codes are prefix codes though, determining the length of the exponent encoding is straightforward.

One starts at the third bit and, including it, counts the number of identical bits that follow. If there is a sequence of N identical bits (whether 0s or 1s) starting from the third bit, then one can deduce that the exponent was encoded on 2N+1 bits.

An example best illustrates this.

1011100110100000000010100000011000101011001.

Starting from the third bit, there is a sequence of three 1s, so the exponent is on 7 bits

10 \textbf{1110011} 0100000000010100000011000101011001.

The next step is to flip all the bits in the exponent encoding if the leading bit is a 0. In the example, no change is needed.

The exponent is then decoded as a modified gamma code (Section \ref{section-gamma-code}) and offset by -2:

\begin{enumerate}
\item The first N+1 bits are replaced with a 1 (in the example: 1011).
\item The obtained bit sequence is decoded as a natural binary representation (11).
\item One substracts 2 (example: 9).
\end{enumerate}

If the obtained exponent is 0, but the exponent sign was encoded as negative (i.e., the overall decimalInfinite encoding begins with 10011 or 00100), an error is raised.

\subsection{Decoding the significand}

The significand is decoded in groups of 10 bits (except the first group which has 4 bits). Each group is decoded as a natural binary representation. The first group gives the digit before the decimal point, the other groups give the digits (three per group) after the decimal point.

If the first group does not deliver a number comprised between 0 and 9, or a subsequent group does not deliver a number comprised between 0 and 999, an error is raised.

Finally, if the overall sign is negative, the complement to 10 is taken instead. An error is raised if the result is not comprised between 1 (included) and 10 (excluded).

\subsection{Summary of decoding errors}
The following errors can be raised upon decoding an invalid sequence:
\begin{itemize}
\item the sequence begins with 01 or 11, but does not correspond to -INF, +INF or NaN (Section \ref{section-fine-tune})
\item 0 was encoded as a negative exponent. (an encoded sequence cannot begin with 10011 or 00100)
\item a tetrade or a declet is outside of the [0,9] or [000,999] range.
\item the overall significand, after possibly taking the complement to 10, is outside of the [1, 10) range, that is:
\begin{itemize}
\item the encoded tetrade is 0 for a positive decimal, or
\item there are non-zero encoded declets after a 9 for a negative decimal.
\end{itemize}
\end{itemize}

\section{Why it is order-preserving}
\label{section-proof}

The encoding is designed in such a way that, if $a < b$ (we treat zero separately), then the encoded $a$ ($STEM_a$) comes lexicographically before ($<<$) the encoded $b$ ($STEM_b$).

A proof thereof now follows.

We index s, t, e, m, S, T, E and M with a and b, that is, $a$'s (absolute) exponent is called $e_a$, $b$'s exponent is called $e_b$. $a$'s significand is called $m_a$, b's significand is called $m_b$, and so on.
 
\begin{enumerate}
\item If $a$ is negative and $b$ is positive, then $S_a=00$ and $S_b=10$, so that $STEM_a << STEM_b$.
\item If $a$ and $b$ are both positive, then $S_a=S_b=10$.
\begin{enumerate}
  \item If $a$'s exponent is negative and $b$'s exponent is non-negative, the next digit in $STEM_a$ ($T_a$) will be a 0 and that of $STEM_b$ ($T_b$) a 1, so that $STEM_a<< STEM_b$.

  Otherwise the exponents have the same sign.
  \item If $a$'s exponent and $b$'s exponent are both non-negative ($T_a=T_b=1$), and $e_a+2$ has less digits than $e_b+2$, then $TE_a$ will have less 1s than $TE_b$ in front of the next 0, so that $STEM_a << STEM_b$.
  \item If $a$'s exponent and $b$'s exponent are both negative ($T_a=T_b=0$), and $e_a+2$ has more digits than $e_b+2$, then $TE_a$ will have more 0s than $TE_b$ in front of the next 1, so that $STEM_a << STEM_b$.

  Otherwise the exponents have the same sign and their offsets by 2 have the same number of bits.
  \item If $a$'s exponent and $b$'s exponent are both non-negative ($T_a=T_b=1$) but different ($e_a < e_b$) and $e_a+2$ has as many digits ($N$) as $e_b+2$, then $TE_a$ and $TE_b$ will both have $N-1$ 1s followed by a 0. The next $N-1$ digits after the 0 in $TE_a$ and $TE_b$ correspond to a natural binary representation (with no leading 1) of $e_a$ and $e_b$ respectively, so that $STEM_a << STEM_b$ because the natural binary representations preserve order given a fix number of digits.
  \item If $a$'s exponent and $b$'s exponent are both negative ($T_a=T_b=0$) but different ($e_a > e_b$) and $e_a+2$ has as many digits ($N$) than $e_b+2$, then $TE_a$ and $TE_b$ will both have $N-1$ 0s followed by a 1. The next $N-1$ digits after the 0 in $TE_a$ and $TE_b$ correspond to an inverted natural binary representation, with no leading 0, of $e_a$ and $e_b$ respectively. Since $e_a > e_b$, $E_A << E_B$ because it's inverted, and $STEM_a << STEM_b$.

  Otherwise the exponents are identical.
  \item If $a$'s exponent and $b$'s exponent are equal, then $m_a < m_b$ and then $TE_a=TE_b$. $M_a$ and $M_b$, organized in one group of 4, then groups of 10, are all natural binary representations of the symbols of $m_a$ and $m_b$ in base 1000, and preserve the order, so that $M_a<<M_b$ and $STEM_a << STEM_b$.
  \end{enumerate}

\item If $a$ and $b$ are both negative, then $S_a=S_b=00$.
  \begin{enumerate}
  \item If $a$'s exponent is non-negative and $b$'s exponent is negative, the next digit in $STEM_a$ will be a 0 and that of $STEM_b$ a 1, so that $STEM_a << STEM_b$.

  Otherwise the exponents have the same sign.
  \item If $a$'s exponent and $b$'s exponent are both non-negative ($T_a=T_b=0$), and $e_a+2$ has more digits than $e_b+2$, then $TE_a$ will have more 0s than $TE_b$ in front of the next 1, so that $STEM_a << STEM_b$.
  \item If $a$'s exponent and $b$'s exponent are both negative ($T_a=T_b=1$), and $e_a+2$ has less digits than $e_b+2$, then $TE_a$ will have less 1s than $TE_b$ in front of the next 0, so that $STEM_a << STEM_b$.

  Otherwise the exponents have the same sign and their offsets by 2 have the same number of bits.
  \item If $a$'s exponent and $b$'s exponent are both negative ($T_a=T_b=1$) but different ($e_a < e_b$) and $e_a+2$ has as many digits ($N$) than $e_b+2$, then $TE_a$ and $TE_b$ will both have $N-1$ 1s followed by a 0. The next $N-1$ digits after the 0 in $TE_a$ and $TE_b$ correspond to a natural binary representation (with no leading 1) of $e_a$ and $e_b$ respectively, so that $STEM_a << STEM_b$ because the natural binary representations preserve order given a fix number of digits.
  \item If $a$'s exponent and $b$'s exponent are both non-negative ($T_a=T_b=0$) but different ($e_a > e_b$) and $e_a+2$ has as many digits ($N$) than $e_b+2$, then $TE_a$ and $TE_b$ will both have $N-1$ 0s followed by a 1. The next $N-1$ digits after the 0 in $TE_a$ and $TE_b$ correspond to an inverted natural binary representation, with no leading 0, of $c$ and $d$ respectively. Since $e_a > e_b$, $E_A << E_B$ because it's inverted, and $STEM_a << STEM_b$.

  Otherwise the exponents are identical.
  \item If $a$'s exponent and $b$'s exponent are equal, then $m_a > m_b$ and $TE_a=TE_b$. $M_a$ and $M_b$, organized in one group of 4, then groups of 10, are all natural binary representations of the symbols of $10-m_a$ and $10-m_b$ in base 1000, and preserve the order. Since $10-m_a < 10-m_b$, $M_a<<M_b$ and and $STEM_a << STEM_b$.
  \end{enumerate}
\item 0 is encoded as 10 and 10 is lexicographically smaller than the encodings of negative decimals, which begin with 00. it is lexicographically greater than the encodings of positive decimals, which begin with 10 followed by at least one further digit.
\end{enumerate}

\section{Examples}
\label{section-examples}

Figure \ref{figure-first-integers} shows the decimalInfinite encoding for the smallest integers. Integers with an absolute value between 1 and 10 are encoded on 9 bits, and until 100 on 19 bits (less if trailing zeros are removed, as shown in Section \ref{section-trailing-zeros}).

\begin{figure}[h]
\caption{The encoding of the smallest integers (the space separators are only here for an easier read).}
\label{figure-first-integers}
\center
\begin{tabular}{|l|l|}
\hline
Decimal & decimalInfinite encoding \\
\hline
-15 & 00 010 1000 0111110100\\
\hline
-14 & 00 010 1000 1011111000\\
\hline
-13 & 00 010 1000 1010111100\\
\hline
-12 & 00 010 1000 1100100000\\
\hline
-11 & 00 010 1000 1110000100\\
\hline
-10 & 00 010 1001\\
\hline
-9 & 00 011 0010\\
\hline
-8 & 00 011 0010\\
\hline
-7 & 00 011 0011\\
\hline
-6 & 00 011 0100\\
\hline
-5 & 00 011 0101\\
\hline
-4 & 00 011 0110\\
\hline
-3 & 00 011 0111\\
\hline
-2 & 00 011 1000\\
\hline
-1 & 00 011 1001\\
\hline
0 & 10 \\
\hline
1 & 10 100 0001\\
\hline
2 & 10 100 0010\\
\hline
3 & 10 100 0011\\
\hline
4 & 10 100 0100\\
\hline
5 & 10 100 0101\\
\hline
6 & 10 100 0110\\
\hline
7 & 10 100 0111\\
\hline
8 & 10 100 1000\\
\hline
9 & 10 100 1001\\
\hline
10 & 10 101 0001\\
\hline
11 & 10 101 0001 0001100100\\
\hline
12 & 10 101 0001 0011001000\\
\hline
13 & 10 101 0001 0100101100\\
\hline
14 & 10 101 0001 0110010000\\
\hline
15 & 10 101 0001 0111110100\\
\hline
\hline
\end{tabular}
\end{figure}

\section{Fine-tuning the scheme}
\label{section-fine-tune}

\subsection{Special numbers}

Special numbers such as positive and negative infinity, negative zero can also be encoded in such a way that the lexicographic order still holds, as shown on figure \ref{figure-sign-extended}. NaN can also be encoded (even if the order does not apply in this case).

\begin{figure}
\caption{Adding special numbers.}
\label{figure-sign-extended}
\center
\begin{tabular}{|l|l|}
\hline
00 & -INF \\
\hline
00... &  negative sign (e.g., $-4.3\times10^3$)\\
\hline
01 & negative zero \\
\hline
10 & positive zero \\
\hline
10... & positive sign (e.g., $4.3\times10^3$)\\
\hline
11 & +INF \\
\hline
111 & NaN \\
\hline
\end{tabular}
\end{figure}

\subsection{Trailing zeros}
\label{section-trailing-zeros}

To save space, trailing zeros can be removed from the binary encoding and added back while decoding (to fit the size of the last declet group).

\subsection{Fix-length variant}

In environments where encoding preserving lexicographic order is not supported across different lengths (this is the case with MongoDB's ordering of binaries), this encoding can be adapted to work at the cost of limiting the range.
A prefix of the binary encoding can be taken as an approximation of the encoded decimal, possibly padded with leading 0s if too short. This works as long as the total stored lengths exceeds the length of the encoding of the sign and exponent, which limits the range.

\subsection{Prefix code variant}
\label{section-prefix-code}

As suggested by Nathan Hurst, decimalInfinite can be turned into a prefix code (self-delimiting). This can be achieved by adding a bit after each tetrade and declet in the significand encoding, to indicate whether further bits follow (1) or not (0), as is done in the Rishe encoding.

\section{Complexity}
\label{section-complexity}
The storage space is linear in the size (number of digits) of the significand, and logarithmic in the exponent, that is, double-logarithmic in the decimal.

The encoding size of a decimal can be computed with:

$$5 + 2 \lfloor\log_2 (e+2)\rfloor + \frac{10}{3} (|m|-1)$$

Where $|m|$ denotes the number of decimal digits in the significand. Asymptotically, it can be approximated further with $$2 \log_2 e + \frac{10}{3}|m|$$ that is $$O(\log e + |m|)$$

The encoding of the significand uses a common approach that is very compact in terms of entropy, and making it more compact (for example, by grouping bits in bigger groups) would increase computational complexity. The factor of $\frac{10}{3}$ is slightly better than the factor of 4 specified in \cite{Rishe1992}, where 8 bits are on average necessary for each additional two significant digits. For the sake of a fair comparison, note that the prefix-code version (Section \ref{section-prefix-code}) of decimalInfinite would increase the factor to $\frac{11}{3}$. Also, the Rishe encoding of very small integers is, by design, more compact.

The encoding of the exponent deviates from an optimal size by a constant factor of 2, which is the cost of using the Gamma prefix code. For large decimals, the Rishe encoding contributes a factor of $0.8 \log_{10} d$ (logarithmic) to the exponent part (semi-progressive intervals require 8 bits for an increase of 10 in the exponent), whereas decimalInfinite contributes only $2 \log_{10}\log_{10} d$ (double-logari\-thmic).

If only non-negative integers ($i$) are considered, then $$|m|= \log_{10} i$$ is asymptotically more significant than $$\log_2 e=\log_2\lfloor\log_{10} i\rfloor$$ so that the encoding size grows logarithmically in $i$, which is an optimal complexity (see Figure \ref{figure-size}).

This logarithmic complexity is identical to that of the Rishe encoding, except that the constant is slightly lower (3.3 vs. 4.8).

\begin{figure}
\label{figure-size}
\caption{Asymptotic size of the encoding for integers (obtained experimentally with the JSONiq implementation).}
The blue lines show the fix-length binary natural representation on 32, 64 and 128 bits on the range they cover.

\begin{tikzpicture}
\begin{axis}[
  xlabel=Integer (logarithmic scale),
  ylabel=Size of the encoding (in bits),
  xmode=log,
  xmin=1,
  ymin=0
]
\addplot[sharp plot] table {integer_size.dat};
\addplot[blue,domain=1:4e9]{32};
\addplot[blue,domain=1:1e19]{64};
\addplot[blue,domain=1:3e38]{128};
\end{axis}
\end{tikzpicture}
\end{figure}

\section{Implementation}
\label{section-implementation}
The encoding and decoding algorithms were successfully implemented in C++, and are used to store JSON numbers lossless in BSON binaries on an underlying MongoDB layer. The decimals are stored with a special library supporting big numbers, however for efficiency reasons we do use classical limited-range integers when dealing with exponents or the number of digits of the decimal to encode. This is not a limitation of the decimalInfinite algorithm, but more a pragmatic decision that it is unlikely that decimals on the order of magnitude of $10^{2^{32}}$, or with a precision of $2^{32}$ digits would be encountered in a real-life setting.

Unfortunately, MongoDB does not use a full lexicographic ordering of binaries (rather, pseudo-lexicographically), such that padding to a fixed length was needed for this vendor.

The code has been running with no known issues on production servers since 2012.

Also, an implementation in JSONiq is available on github \cite{decimalInfiniteJSONiq}, and is exposed publicly via a very basic REST API.

\section{Conclusion}

We introduced a binary encoding that supports the entire decimal value space including special numbers, that does not lose precision, and that preserves the order, in the sense that the encoding is a homomorphism between the decimals (sorted naturally) and the bit sequences (sorted fully lexicographically). This encoding is simple to explain: its specification in this paper fits on 2 pages. It is not parameterized in any way, making it implementable and understandable in a straightforward way.

\section{Acknowledgments}
decimalInfinite was implemented on the 28.io platform with the help of Matthias Brantner and Till Westmann. Also, many thanks to Nathan Hurst for his prompt and useful feedback.

\bibliographystyle{abbrv}

\end{document}